\documentclass[a4paper,11pt]{article}
\usepackage{pos}
\usepackage{graphicx}
\usepackage{color}
\usepackage{float}
\usepackage{subfig}
\usepackage{ragged2e}      
\usepackage[export]{adjustbox}
\usepackage{multirow}
\usepackage{array}
\usepackage{blindtext}
\usepackage{multicol}
\usepackage{color, colortbl}
\usepackage[export]{adjustbox}
\usepackage{changepage}
\usepackage{tikz}

\title{Two-Loop Amplitude Reduction with \textsc{HELAC}}

\author[a]{Giuseppe Bevilacqua}
\author*[a,b]{Dhimiter Canko}
\author[a]{Costas Papadopoulos}

\affiliation[a]{Institute of Nuclear and Particle Physics, NCSR ”Demokritos”,\\
Agia Paraskevi 15310, Greece}

\affiliation[b]{Department of Physics, University of Athens,\\
Zografou 15784, Greece}

\emailAdd{bevilacqua@inp.demokritos.gr}
\emailAdd{jimcanko@phys.uoa.gr}
\emailAdd{costas.papadopoulos@cern.ch}

\abstract{ We discuss recent progress towards extending the \textsc{Helac} framework to the calculation of  two-loop amplitudes. A general algorithm for the automated computation of two-loop integrands is described. The algorithm covers all the steps of the computation, from the generation of loop topologies up to the construction  of recursion relations for two-loop integrands. Finally, first steps towards the formulation of a new approach for reducing two-loop amplitudes to a basis of master integrals are discussed.}

\FullConference{16th International Symposium on Radiative Corrections: Applications of Quantum Field Theory to Phenomenology (RADCOR2023)\\
28th May - 2nd June, 2023\\
Crieff, Scotland, UK\\}

\begin{document}
\maketitle

\section{Introduction}
Continuous improvements in the precision of the data collected at the LHC are demanding corresponding improvements of theoretical predictions.
This is imperative for QCD processes in particular, given that hadronic interactions are ubiquitous at the LHC. The calculation of hard-scattering cross sections is at the core of perturbative QCD and is organised in terms of an expansion in powers of the strong coupling constant. On a general ground, the precision frontier of perturbative calculations to date is set to NNLO\footnote{For $2 \to 2$  processes the current frontier is N$^3$LO.}. An up-to-date reference list of processes which are desired with the highest accuracy possible can be found in Ref.\cite{Huss:2022ful}.

Generally speaking, the bottleneck of NNLO calculations is identified in double-virtual corrections, more precisely in the calculation of two-loop amplitudes. Although the workflow of two-loop calculations varies upon the specific method considered, the key aspects can be summarized as follows:
\begin{enumerate}
\item \textit{construction} of two-loop integrands. This can be achieved either computing individual Feynman diagrams or using recursion relations;
\item \textit{reduction} of two-loop amplitudes  in terms of Master Integrals \cite{Ita:2015tya, Mastrolia:2016dhn, Abreu:2020xvt, Badger:2022ncb, Badger:2021ega, Badger:2021nhg, Hartanto:2019uvl, Abreu:2021asb, Badger:2021imn, Chawdhry:2021mkw, Agarwal:2021grm, Chawdhry:2020for, Abreu:2020cwb, Abreu:2021oya, Abreu:2019odu, Abreu:2018zmy, Abreu:2023bdp, Agarwal:2021vdh, Peraro:2016wsq, Pozzorini:2022ohr}. This can be achieved at integrand level or at integral level; 
\item  \textit{calculation} of Master Integrals \cite{DE1, DE2, Gehrmann:1999as, Henn:2013pwa, Papadopoulos:2014lla, Papadopoulos:2015jft, Liu:2017jxz, Gehrmann:2018yef, Chicherin:2018old, Moriello:2019yhu, Chicherin:2020oor, Abreu:2020jxa, Canko:2020ylt, Abreu:2021smk, Kardos:2022tpo, Chicherin:2021dyp, Abreu:2023rco, SecDec, Borinsky:2020rqs} based on analytical or (semi-)numerical methods.
\end{enumerate}
Each of these steps comes with its own challenges, and the developements during the last years have yielded a number of results for $2 \to 3$ cross sections at NNLO accuracy \cite{Badger:2023mgf, Kallweit:2020gcp, Chawdhry:2021hkp, Czakon:2021mjy, Hartanto:2022qhh}.

In this contribution we present steps towards the construction of \textsc{Helac-2loop}, a framework for automated two-loop calculations. We focus in particular on items 1 and 2 in the list above. In Section \ref{Sec:integrands} we illustrate the basic details of the algorithm for the computation of two-loop integrand functions. A few benchmark results are presented in Section \ref{Sec:benchmarks}. Finally, in Section \ref{Sec:reduction} we sketch the approach for the reduction of two-loop amplitudes that we plan to implement in our framework. 

\section{Construction of two-loop integrands}
\label{Sec:integrands}

Before starting the discussion, it is convenient to introduce some basic notation. 
We denote a generic scattering process described by $n$ external particles with the flavors of the latter, $\{f_1,f_2,\dots,f_n\}$. Color degrees of freedom are treated in the \textit{color-flow} representation \cite{tHooft:1973alw}. In this representation external gluons are labelled with a doublet of color indices $\{i,j\}$, while external quarks and anti-quarks receive the labels $\{i,0\}$ and $\{0,j\}$ respectively. A generic color state is uniquely identified by the product $\delta^{i_{\sigma_1}}_{j_1}\delta^{i_{\sigma_2}}_{j_2}\cdots\delta^{i_{\sigma_n}}_{j_n}$, where $\{\sigma_1,\sigma_2,\dots,\sigma_n\}$ denotes a permutation of the set $\{1,2,\dots,n\}$. Thus, for a generic process consisting of $n_g$ external gluons and $n_q + n_{\bar{q}}$ quarks, the number of color states is $(n_g + (n_q + n_{\bar{q}})/2)!$ (see \textit{e.g.} Ref. \cite{Cafarella:2007pc, Kanaki:2000ey}
 for more details).
The calculation of scattering amplitudes is organised recursively. The building blocks of the recursion are objects named currents, \textit{i.e.} tree-level sub-amplitudes built out of a subset $S_B \subset \{1,\dots,n\}$ of the external particles. Currents are combined into sub-amplitudes of increasing complexity till the full amplitude is reached (see Ref. \cite{Cafarella:2007pc, Kanaki:2000ey}
 for more details). Each current (hereafter also denoted \textit{blob}, or $B$) is labeled with a unique integer  $\rm{ID}(B) = \sum_{i \in S_B}2^{i-1}$. 

\begin{figure}
\hspace{0.cm}
\includegraphics[clip, trim=2.cm 13cm 2.cm 12.5cm, width=0.99\textwidth]{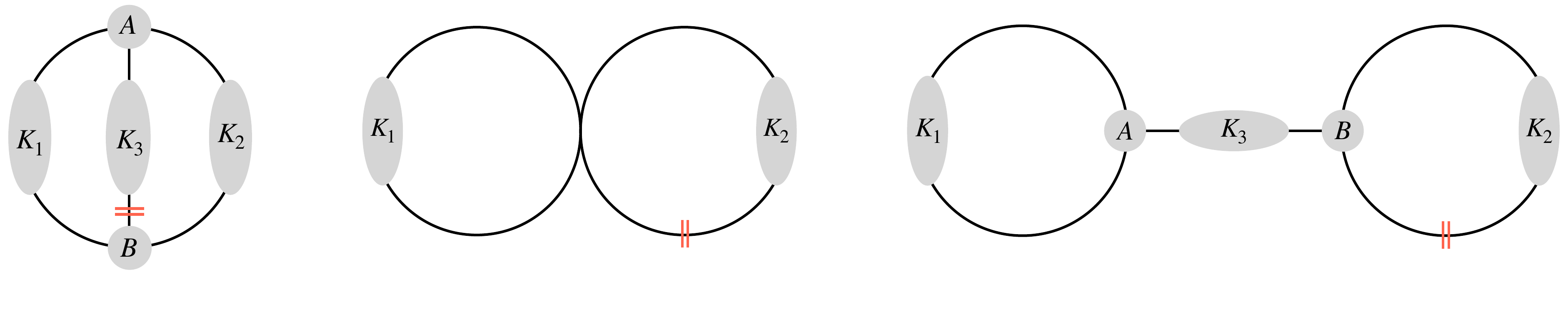}
\put(-399,-5){ \small{(a)} }
\put(-276,-5){ \small{(b)} }
\put(-98,-5){ \small{(c)} }
\caption{Master categories of two-loop topologies: \textit{Theta} (a), \textit{Infinity} (b),  \textit{Dumbbell} (c). The red lines indicate the propagator that is cut in order to express the two-loop topology as an equivalent $(n+2)$-particle process at one-loop (explained in the text).}
\label{Fig:theta_infinity_dumbbell}
\end{figure}

The first step towards the construction of two-loop integrands consists in the generation of \textit{loop topologies}. A key observation is that all  two-loop topologies describing arbitrary processes in the Standard Model\footnote{More generally, this statement is true for any model whose  Feynman rules include up to four-particle vertices} (SM) fall into one of the three master categories shown in Figure \ref{Fig:theta_infinity_dumbbell}, that we name "\textit{Theta}", "\textit{Infinity}" and "\textit{Dumbbell}" for brevity.
The shaded-gray areas $K_i$ appearing therein, that we also refer to as \textit{sectors}, denote schematically a set $\{B_1,\dots,B_{L_i}\}$ of tree-level blobs inserting to the $i$-th propagator of the loop, where $K_i = \sum_{j=1}^{L_i}\rm{ID}(B_j)$. Similarly $A,B$ denote generically a blob that can possibly insert into each of the loop vertices. The number of blobs appearing in each sector, together with the ordering in which they are displaced, defines a topology. We note, however, that there is no one-to-one correspondence between blob orderings and loop topologies: the reason is that the latter are symmetric under the operations of left/-right and up-down reversal. Furthermore, \textit{Theta} topologies are symmetric under any permutation of the sectors $K_1,K_2,K_3$. Configurations of blob orderings that are left-right or up-down symmetric are identified at this stage in order to avoid double counting. For the same reason, all \textit{Theta} topologies are generated according to the rule $K_1>K_2>K_3$. Further details concerning the topological generation phase can be found in Ref. \cite{Bevilacqua:2021dym}.

\begin{figure}
\hspace{-0.7cm}
\includegraphics[clip, trim=5.cm 7cm 7.cm 8cm, width=0.5\textwidth]{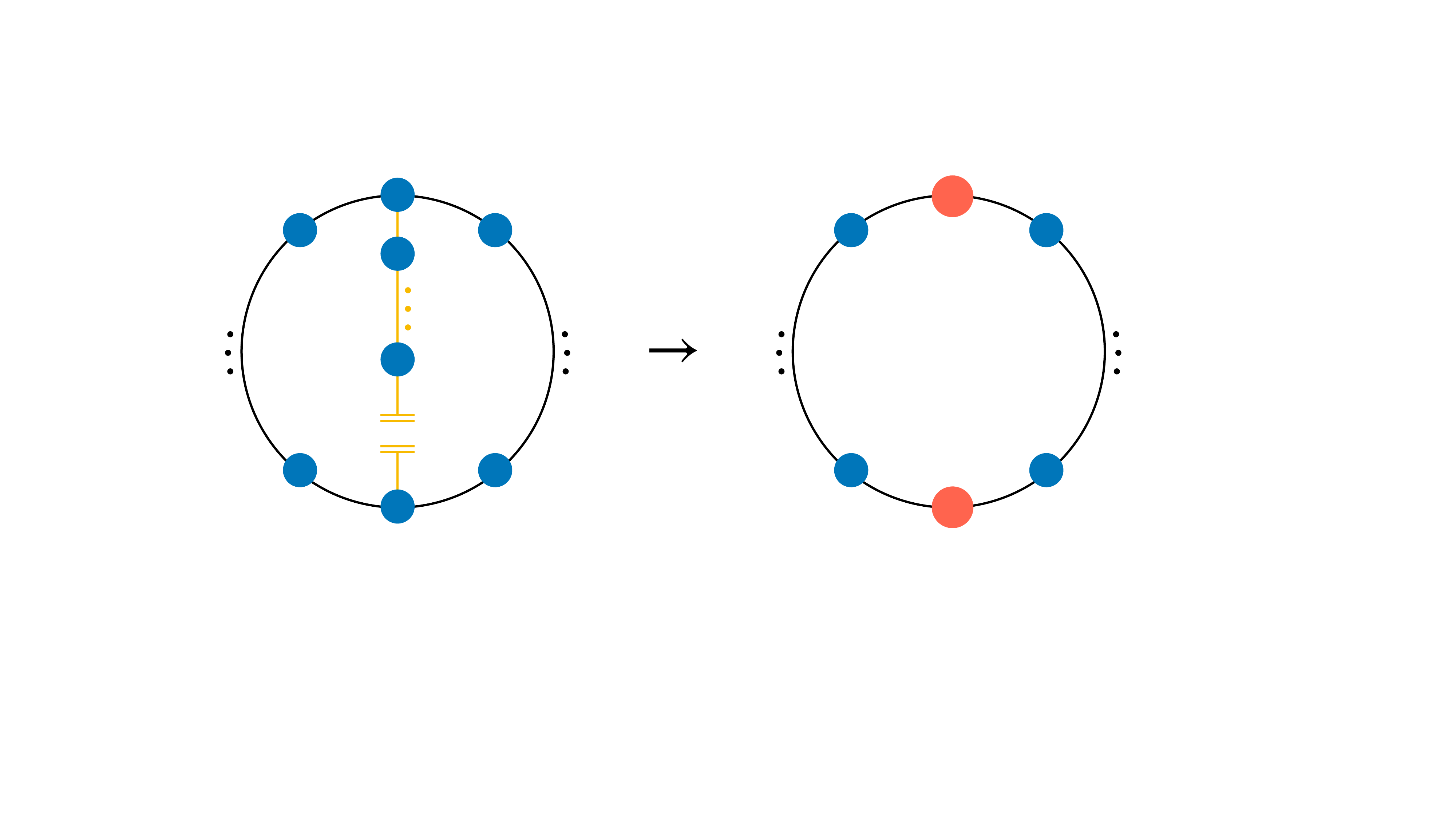}
\put(15,-5){ \includegraphics[clip, trim=7.cm 7cm 7.cm 10cm, width=0.55\textwidth]{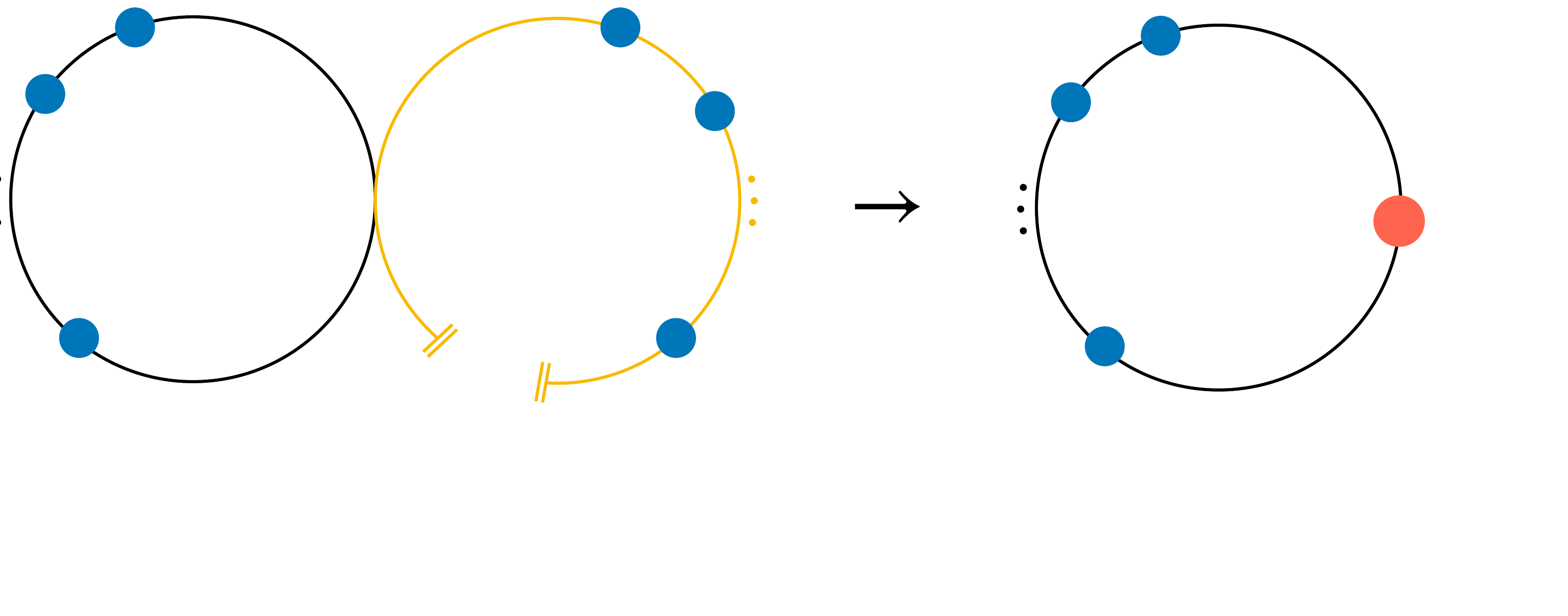} }
\put(-100,-70){ \includegraphics[clip, trim=2.5cm 11cm 5.cm 10cm, width=0.6\textwidth]{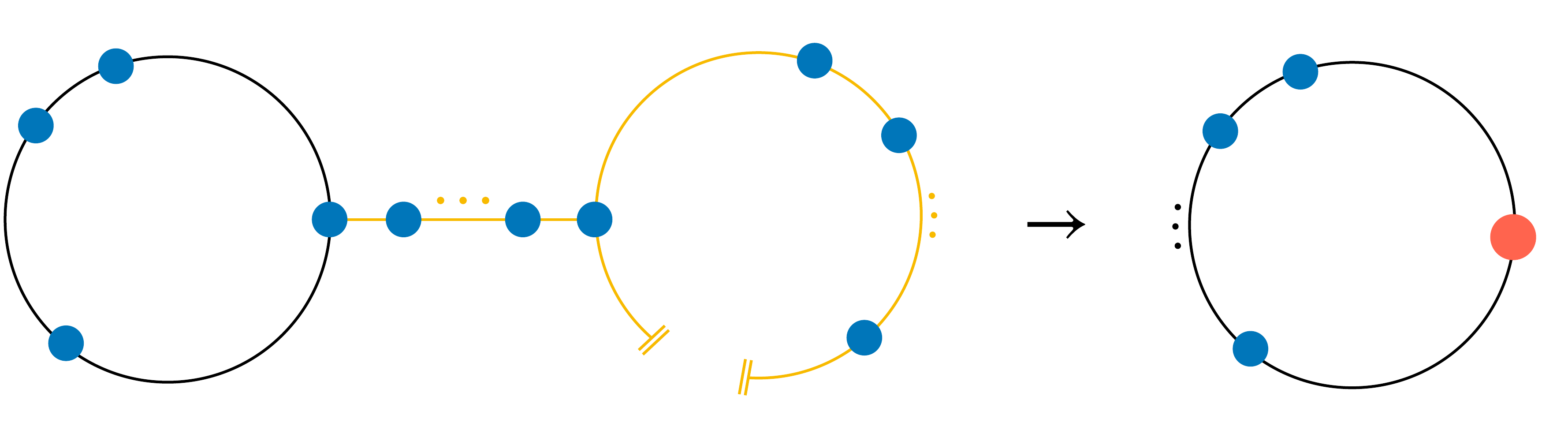} }
\put(-127,20){ \small{(a)} }
\put(140,20){ \small{(b)} }
\put(65,-70){ \small{(c)} }
\caption{Relations among cut two-loop topologies and their equivalent one-loop configuration in $(n+2)$ kinematics: \textit{Theta} (a), \textit{Infinity} (b), \textit{Dumbbell} (c). The blobs denote tree-level currents which are attached to the loop propagators.}
\label{Fig:oneloop}
\end{figure}

\begin{figure}
\includegraphics[clip, trim=0cm 22cm 0cm 0cm, width=0.95\textwidth]{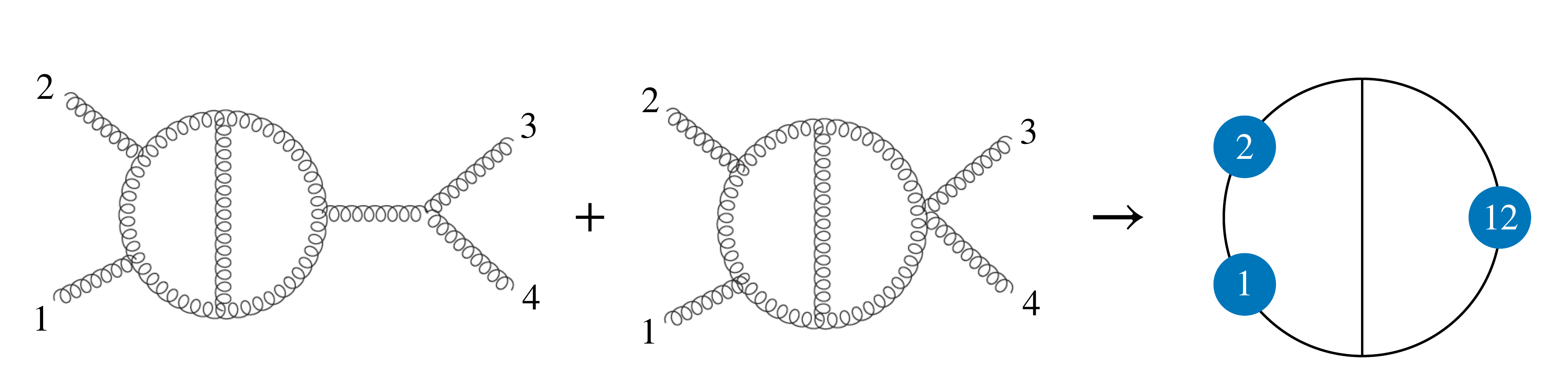} \\
\includegraphics[width=0.95\textwidth]{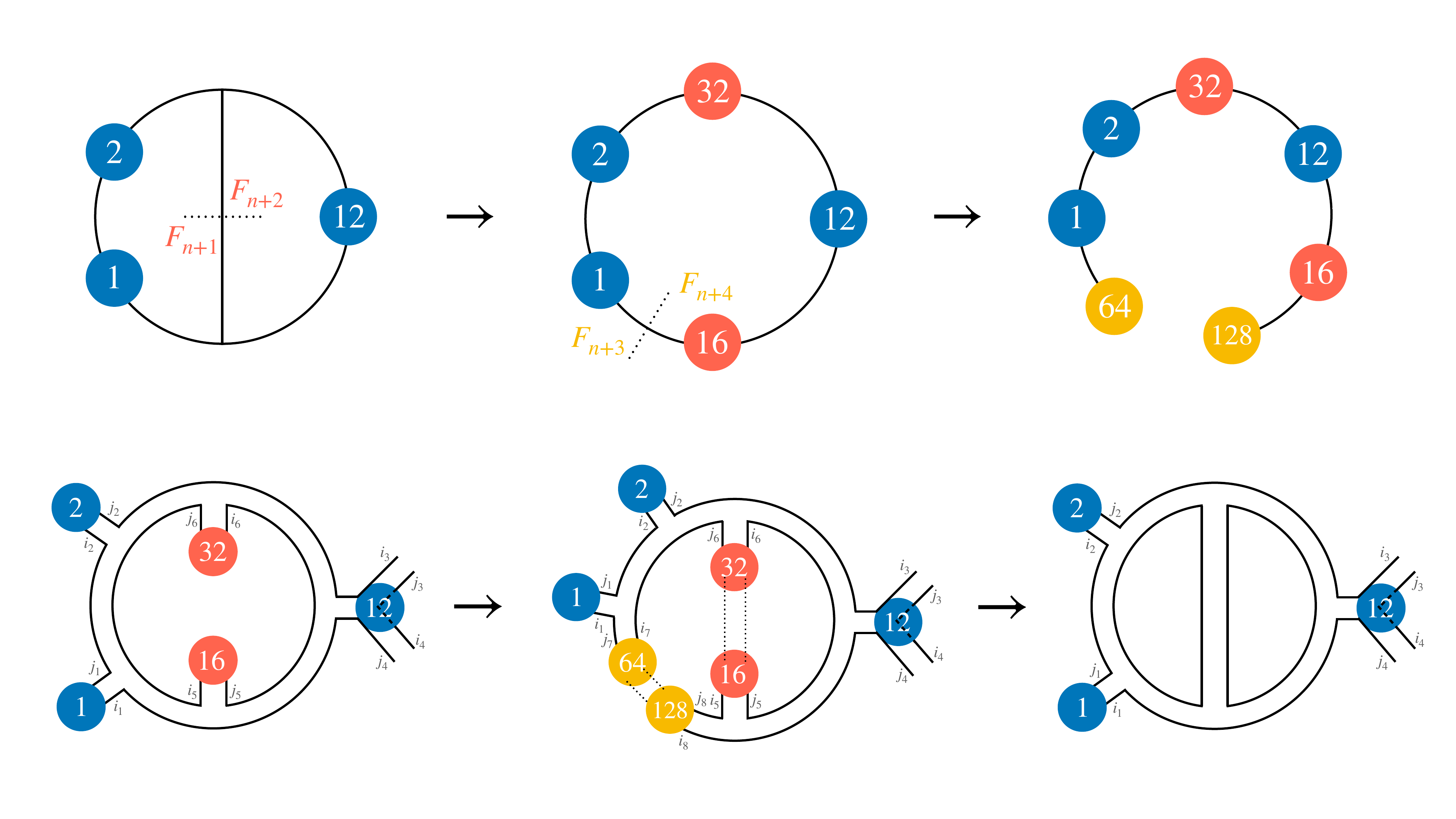}
\scriptsize{
\put(-384,4){ $\delta^{i_2}_{j_1}\delta^{i_3}_{j_2}\delta^{i_4}_{j_3}\delta^{i_1}_{j_4}\delta^{i_6}_{j_5}\delta^{i_5}_{j_6}$ }
\put(-250,4){ $\delta^{i_2}_{j_1}\delta^{i_3}_{j_2}\delta^{i_4}_{j_3}\delta^{i_8}_{j_4}\delta^{i_6}_{j_5}\delta^{i_7}_{j_6}\delta^{i_1}_{j_7}\delta^{i_5}_{j_8}$ }
\put(-102,4){ $N_c^2 \cdot \delta^{i_2}_{j_1}\delta^{i_3}_{j_2}\delta^{i_4}_{j_3}\delta^{i_1}_{j_4}$ }
}
\caption{Schematic illustration of color-flavor dressing for a simple two-loop topology.}
\label{Fig:numerator_example}
\end{figure}

The topologies generated at this stage carry no information about flavor and color of the propagators in the loop. This information is generated in the next step, referred as \textit{color-flavor dressing} (schematically illustrated in Figure \ref{Fig:numerator_example} in the context of a simple example). The procedure can be summarised in the following steps:
\begin{enumerate}
\item we cut one propagator in the two-loop topology (using the conventions shown in Figure \ref{Fig:theta_infinity_dumbbell}). In this way we take the perspective of a one-loop topology describing a $(n+2)$-particle process with flavors $\{f_1,f_2,\dots,f_n,F_{n+1},F_{n+2}\}$. We refer to the latter as the \textit{equivalent one-loop configuration} for brevity. The flavors $F_{n+1},F_{n+2}$ refer to the two additional particles originated by the cut and can take any value allowed to run into the loop. This step is schematised in Figure \ref{Fig:oneloop}. 
We draw blobs using two different colors to stress that they are computed using different approaches. The blue blobs denote tree-level currents computed with standard Dyson-Schwinger recursion, \textit{i.e.} all possible sub-amplitudes compatible with the external-particle content are incorporated. By contrast, the recursion defining the red blobs is partially constrained according to the non-trivial internal structure induced by the loop propagator that has been cut (shown in yellow);
\item we treat $F_{n+1},F_{n+2}$ as external particles with truly independent degrees of freedom in the context of the one-loop process $\{f_1,f_2,\dots,f_n,F_{n+1},F_{n+2}\}$. This allows us to reuse (with proper modifications) the apparatus for one-loop integrand computations already developed in \textsc{Helac-1loop} \cite{vanHameren:2009dr, Bevilacqua:2011xh}
. Of course, $F_{n+1},F_{n+2}$ are constrained in both flavor and color as they are originated by cutting a propagator. Thus, in our strategy, a redundant number of degrees of freedom is considered at this intermediate stage. The latter shall be projected into the lower-dimensional space describing the physical $n$-particle process\footnote{For the color degrees of freedom, the projection is understood as a contraction of the form $\left( \delta^{i_{\sigma_1}}_{j_1}\delta^{i_{\sigma_2}}_{j_2}\cdots\delta^{i_{\sigma_n}}_{j_n}\delta^{i_{\sigma_{n+1}}}_{j_{n+1}}\delta^{i_{\sigma_{n+2}}}_{j_{n+2}} \right) \left( \delta^{j_{n+1}}_{i_{n+2}}\delta^{j_{n+2}}_{i_{n+1}} \right) = (N_C)^{\alpha} \cdot \delta^{i_{\sigma^{\prime}_1}}_{j_1}\delta^{i_{\sigma^{\prime}_2}}_{j_2}\cdots\delta^{i_{\sigma^{\prime}_n}}_{j_n}$, where $\{\sigma^{\prime}_1,\dots,\sigma^{\prime}_n\}$ ($\{\sigma_1,\dots,\sigma_{n+2}\}$) is a permutation of $\{1,\dots,n\}$ ($\{1,\dots,n+2\}$), $N_C$ is the number of colors, and $\alpha=0 \, \text{or} \, 1$ depending on the configuration.};
\item with reference to the one-loop configuration mentioned in the previous step, the construction of the integrand proceeds following the conceptual design of \textsc{Helac-1loop}. The one-loop propagators are dressed with colors using all possible configurations which conserve color flow at each vertex of the loop. Similarly, propagators are dressed with flavors in all ways that are compatible with the SM. The one-loop topology, in turn, is cut, originating two additional external particles $F_{n+3},F_{n+4}$. This sets the stage of a tree-level, $(n+4)$-particle process with flavors $\{f_1,f_2,\dots,f_n,F_{n+1},F_{n+2},F_{n+3},F_{n+4}\}$. At this point the computation of one-loop numerator functions can be addressed with fully automated tree-level technology. More technical details can be found in  Ref. \cite{vanHameren:2009dr, Bevilacqua:2011xh}
.
\end{enumerate}

\begin{figure}[h!]
\hspace{0.8cm}
\includegraphics[width=0.86\textwidth]{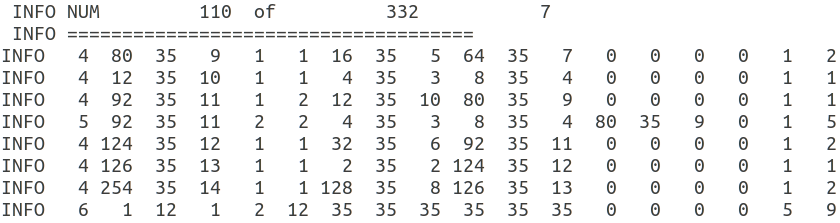}
\caption{Internal representation of a two-loop numerator in the \textsc{Helac} framework. All lines except the last one are in the same format as in \textsc{Helac-1Loop}~\cite{Bevilacqua:2011xh}. The last line includes in addition information on the two-loop topology~\cite{new_paper}: the second and third numbers, 1 and 12, are used to specify the denominator structure of the given numerator.}
\label{Fig:skeleton_numerator}
\end{figure}

In the last step, a set of recursion relations is constructed for each two-loop topology. These encode the instructions to compute numerators that will be used by \textsc{Helac} during numerical evaluations. They can be generated once and stored in a file named \textit{skeleton}. For illustrative purposes, we provide in Figure \ref{Fig:skeleton_numerator}  an example showing how a two-loop numerator is represented in our framework. The example refers to the 110th numerator out of a total of 332 numerators contributing to a specific color connection stored in the skeleton. Each line reports the instructions to compute a single current involved in the recursive calculation of the numerator. The information concerning the two-loop topology is stored in the last line.

We have performed a number of tests at the level of individual two-loop numerators in order to validate our implementation. Our results have been found in excellent agreement with the numerical output generated with the help of \textsc{FeynArts + FeynCalc} \cite{Hahn:2000kx, Shtabovenko:2020gxv}.

\section{Benchmark results}
\label{Sec:benchmarks}

To give an idea of the computational cost of two-loop numerator generations, we provide in Table \ref{Tab:CPU_mem} some details concerning  the requirements of CPU time and disk occupancy for a few benchmark processes. We focus in particular on four-gluon and five-gluon amplitudes. We consider the Leading Color (LC) approximation as benchmark for our comparisons. For the four-gluon case we also show results based on Full Color (FC). The purpose is to understand how the computational cost scales with increasing multiplicities and/or improved color treatment. In all cases we also report the total number of two-loop numerators as a measure of the complexity of the calculation. Comparing the first two rows of Table \ref{Tab:CPU_mem} we observe that, when going from LC to FC, the memory size increases by one order of magnitude. This is consistent with the larger number of numerators that are generated and stored in FC. Similarly the CPU time scales almost by a factor of 30. A more dramatic impact on scaling is observed, in contrast, when adding one more gluon in the computation while taking the LC viewpoint, as can be inferred by comparing the second and third lines of Table \ref{Tab:CPU_mem}. In this case, CPU time and memory scale approximately by factors 90 and 30 respectively. We would like to stress that there is still room for better organisation of numerators and optimisation of the whole generation chain, which is currently under study. This is expected to improve the performance both in terms of computing efficiency and storage.

\begin{table}
\begin{center}
\begin{tabular}{|c|c|c|c|c|c|}
 \hline 
\textbf{{Process}} & $\mathbf{N_{\mathrm{\mathbf{Loops}}}}$  & \textbf{{Color}} & \textbf{{Numerators}}  & \textbf{{Size [MB]}} & \textbf{{Timing [s]}} \\  [0.1cm]
 \hline \hline
 $gg\to gg$ & 2 & {Full} & 89392 & 111 & 415  \\ [0.1cm]
 \hline 
 $gg\to gg$  & 2 & {Leading} & 4560 & 9  & 15  \\ [0.1cm]
 \hline 
 $gg\to ggg$ & 2 & {Leading}  & 81480 & 300 & 1303 \\ [0.1cm]
 \hline \hline
 $gg\to gg$ & 1 & {Full} & 768 & 0.5 & 2  \\ [0.1cm]
 \hline 
 $gg\to ggg$  & 1 & {Full} & 11496 & 15 & 533  \\ [0.1cm]
 \hline 
\end{tabular}
\end{center}
\caption{ Details of CPU time and disk space required to generate and store all two-loop numerators for a few reference processes. \textit{Leading} refers to leading-color approximation, \textit{Full} denotes full treatment of color degrees of freedom. All numbers have been obtained on an Intel 1.80 GHz processor with the gfortran compiler using the option -O3. Results of one-loop generations are also reported for comparison. }
\label{Tab:CPU_mem}
\end{table}

\section{Steps towards a new approach for two-loop amplitude reduction}
\label{Sec:reduction} 

Let us consider a two-loop topology characterised by $N_P$ loop propagators for a generic process with $n$ external particles. Let  $\{D_1,\dots,D_{N_P}\}$ be the denominators of the loop propagators (we will refer to them as \textit{propagators} in the following for brevity), $\{p_1,\dots,p_n\}$ the four-dimensional momenta of the external particles and $\{k_1,k_2\}$ the loop momenta. The latter, expressed in $d$ space-time dimensions, are decomposed as $k_i = \bar{k}_i + k^{*}_i$ where $\bar{k}_i$ is the four-dimensional and $k^{*}_i$ the extra-dimensional part. The scalar product reads $k_i \cdot k_j = \bar{k}_i  \cdot \bar{k}_j + \mu_{ij}$ ($i=1,2$), where we have defined $\mu_{ij} = k^{*}_i \cdot k^{*}_j$. 

The most general two-loop integrand can be constructed out of the Lorentz scalars
\begin{equation}
p_i \cdot p_j\,,   \qquad   k_i \cdot k_j\,,   \qquad   k_i \cdot p_j\,,     \qquad   k_i \cdot \eta_j\,,
\label{Eq:scalars}
\end{equation}
where $\eta_j$ are \textit{transverse vectors} defined such that $\eta_i \cdot p_j = 0$. The integrand is understood to be a rational function of the form
\begin{equation}
\mathcal{R} = \frac{\mathcal{N}}{\mathcal{D}} \equiv  \frac{\sum_{a} c_{a}(\vec{s},\varepsilon) \, \left( z^{(a)}_1 \right)^{\beta_1}  \cdots \, \left( z^{(a)}_{n_a} \right)^{\beta_{n_{a}}} }{D_1 \cdots D_{N_P} } \,,
\end{equation}
where the $\beta$'s are integers, $\vec{s}$ denotes generically scalar products of the form $p_i \cdot p_j$ and $z^{(a)} \in S = \{k_i \cdot k_j, \; k_i \cdot p_j, \; k_i \cdot \eta_j \}$. Those scalars $z^{(a)}$ which admit a decomposition in terms of linear combinations of propagators $D_i$ cancel with the denominator. Then, the integrand can be recast in the form
\begin{equation}
\mathcal{R} = \sum_{m=0}^{N_P} \sum_{\sigma} \frac{   \sum_{j} \tilde{c}_{j}^{(\sigma)}(\vec{s},\varepsilon) \, \prod_{k=1}^{n_{\rm T+ISP}^{(m)} } \left( \bar{z}^{(\sigma)}_k \right)^{\alpha_k^{(j)}}   }{ D_{\sigma_1} \cdots D_{\sigma_m}  }
\label{Eq:integrand}
\end{equation}
where the $\alpha$'s are integers and $\sigma$ denotes any possible subset of $\{1,\dots,N_P\}$ consisting of $m$ elements. The residual scalar products appearing in the numerator (that we label as $\bar{z}^{(j)}_k$ for clarity) include both the transverse ones (T), $k_i \cdot \eta_j$, as well as what is  known as \textit{irreducible scalar products} (ISP).
Based on Eq.(\ref{Eq:integrand}), one can express the numerator function $\mathcal{N}$ using the following equation:
\begin{equation}
\mathcal{N} = \sum_{m=0}^{N_P}  \sum_{\sigma}  \sum_{j} \tilde{c}_{j}^{(\sigma)}(\vec{s},\varepsilon)  \, \prod_{k=1}^{n_{\rm T+ISP}^{(m)} } \left( \bar{z}^{(\sigma)}_k \right)^{\alpha_k^{(j)}}  \prod_{i \notin \sigma } D_i  \,.
\label{Eq:ansatz}
\end{equation}
Our goal is to find a method to extract numerically the coefficients $\tilde{c}_{j}^{(\sigma)}(\vec{s},\varepsilon)$ at the \textit{integrand} level.
Terms in Eq.(\ref{Eq:ansatz}) which do not depend explicitly on transverse vectors ($\eta_j$) lead to Feynman integrals which can be addressed using integration-by-part (IBP) techniques \cite{Chetyrkin:1981qh, Tkachov:1981wb, Laporta:2000dsw, Smirnov:2019qkx, Klappert:2020nbg}. Scalar products of the form $(k_i \cdot \eta_j)^P$ (where $P$ is a positive integer) can be eliminated, if present, using the following considerations:
\begin{itemize}
\item if $P$ is \textit{odd}, the corresponding terms vanish upon integration; \smallskip
\item if $P$ is \textit{even}, the corresponding terms  can be cast in a form which vanish upon integration using so-called \textit{traceless completion} \cite{Sotnikov:2019onv, Ita:2015tya}. This is equivalent to a change of basis of the form
\end{itemize}
{\footnotesize
\begin{eqnarray}
& & (k_i \cdot \eta_j )\,(k_l \cdot \eta_j )  \longrightarrow (k_i \cdot \eta_j )\,(k_l \cdot \eta_j ) - \frac{\mu_{il}}{d-4} \\
& & (k_{i}\cdot \eta_j)^2 \, (k_{l}\cdot \eta_j)^2 \longrightarrow (k_{i}\cdot \eta_j)^2 \, (k_{l}\cdot \eta_j)^2 - \frac{(k_{i}\cdot \eta_j)^2 \mu_{ll}+ (k_{l}\cdot \eta_j)^2 \mu_{ii}+4(k_{i}\cdot \eta_j) \, (k_{l}\cdot \eta_j) \mu_{il}}{2(d-4)}
\end{eqnarray}
}

Following the paradigm developed for one-loop calculations, and having all the aforementioned in mind, a possible way of implementing the amplitude reduction within \textsc{Helac-2Loop} can be sketched in the following steps:
\begin{enumerate}
\justifying
\item extracting the 4-dimensional part of $\tilde{c}_{j}^{(\sigma)}(\vec{s},\varepsilon)$ using values for the loop-momenta obtained from the cut equations $D_{\sigma_1}=\cdots=D_{\sigma_{m}}=0$ in an OPP-like approach \cite{Ossola:2006us, Ossola:2007ax};
\item finding the polynomial dependence on $\mu_{ij}$ terms (analogously to the treatment of ${\cal R}_1$ terms in OPP reduction);
\item determining the missing $\varepsilon$-dimensional part of the numerator using two-loop rational terms (${\cal R}_2$ terms).
\item reducing the Feynman integrals containing ISP monomials to Master Integrals using IBP reduction.
\end{enumerate}
Notice that all four steps described above are universal. The integrand basis (steps 1 and 2), the ${\cal R}_2$ terms (step 3) and the IBP reduction (step 4) are independent of the specific process under consideration. On the other hand, we expect that the IBP reduction presents technical challenges when applied to certain two-loop topologies with $5$ or more external legs. Moreover, the implementation of ${\cal R}_2$ rational terms (following up Ref. \cite{Pozzorini:2020hkx, Lang:2020nnl, Lang:2021hnw}) requires further study.

\section{Summary}

We have presented an approach for computing two-loop integrands for arbitrary processes in a fully automated way. Furthermore we have sketched a possible strategy for the reduction of two-loop amplitudes to Master Integrals. These results are part of the ongoing efforts to develop \textsc{Helac-2loop}, a framework for automated two-loop amplitude calculation.

\acknowledgments

This work is supported by the Hellenic Foundation for Research and Innovation (H.F.R.I.) under the "2nd Call for H.F.R.I. Research Projects to support Faculty Members \& Researchers" (Project Number: 02674 HOCTools-II).



\begin{thebibliography}{99}

\bibitem{Huss:2022ful}
A.~Huss, J.~Huston, S.~Jones and M.~Pellen,
\href{https://iopscience.iop.org/article/10.1088/1361-6471/acbaec}{J. Phys. G \textbf{50} (2023) no.4, 043001}
\href{https://arxiv.org/abs/2207.02122}{[arXiv:2207.02122 [hep-ph]]}.


\bibitem{Ita:2015tya}
H.~Ita,
\href{https://journals.aps.org/prd/abstract/10.1103/PhysRevD.94.116015}{Phys. Rev. D \textbf{94} (2016) no.11, 116015}
\href{https://arxiv.org/abs/1510.05626}{[arXiv:1510.05626 [hep-th]]}.

\bibitem{Mastrolia:2016dhn}
P.~Mastrolia, T.~Peraro and A.~Primo,
\href{https://link.springer.com/article/10.1007/JHEP08(2016)164}{JHEP \textbf{08} (2016), 164}
\href{https://arxiv.org/abs/1605.03157}{[arXiv:1605.03157 [hep-ph]]}.

\bibitem{Abreu:2020xvt}
S.~Abreu, J.~Dormans, F.~Febres Cordero, H.~Ita, M.~Kraus, B.~Page, E.~Pascual, M.~S.~Ruf and V.~Sotnikov,
\href{https://www.sciencedirect.com/science/article/pii/S0010465521001818?via%3Dihub}{Comput. Phys. Commun. \textbf{267} (2021), 108069}
\href{https://arxiv.org/abs/2009.11957}{[arXiv:2009.11957 [hep-ph]]}.

\bibitem{Badger:2022ncb}
S.~Badger, H.~B.~Hartanto, J.~Kry\'s and S.~Zoia,
\href{https://arxiv.org/abs/2201.04075}{JHEP \textbf{05} (2022), 035}
\href{https://link.springer.com/article/10.1007/JHEP05(2022)035}{[arXiv:2201.04075 [hep-ph]]}.

\bibitem{Badger:2021ega}
S.~Badger, H.~B.~Hartanto, J.~Kry\'s and S.~Zoia,
\href{https://link.springer.com/article/10.1007/JHEP11(2021)012}{JHEP \textbf{11} (2021), 012}
\href{https://arxiv.org/abs/2107.14733}{[arXiv:2107.14733 [hep-ph]]}.

\bibitem{Badger:2021nhg}
S.~Badger, H.~B.~Hartanto and S.~Zoia,
\href{https://journals.aps.org/prl/abstract/10.1103/PhysRevLett.127.012001}{Phys. Rev. Lett. \textbf{127} (2021) no.1, 012001}
\href{https://arxiv.org/abs/2102.02516}{[arXiv:2102.02516 [hep-ph]]}.

\bibitem{Hartanto:2019uvl}
H.~B.~Hartanto, S.~Badger, C.~Br\o{}nnum-Hansen and T.~Peraro,
\href{https://link.springer.com/article/10.1007/JHEP09(2019)119}{JHEP \textbf{09} (2019), 119}
\href{https://arxiv.org/abs/1906.11862}{[arXiv:1906.11862 [hep-ph]]}.

\bibitem{Abreu:2021asb}
S.~Abreu, F.~Febres Cordero, H.~Ita, M.~Klinkert, B.~Page and V.~Sotnikov,
\href{https://link.springer.com/article/10.1007/JHEP04(2022)042}{JHEP \textbf{04} (2022), 042}
\href{https://arxiv.org/abs/2110.07541}{[arXiv:2110.07541 [hep-ph]]}.

\bibitem{Badger:2021imn}
S.~Badger, C.~Br\o{}nnum-Hansen, D.~Chicherin, T.~Gehrmann, H.~B.~Hartanto, J.~Henn, M.~Marcoli, R.~Moodie, T.~Peraro and S.~Zoia,
\href{https://link.springer.com/article/10.1007/JHEP11(2021)083}{JHEP \textbf{11} (2021), 083}
\href{https://arxiv.org/abs/2106.08664}{[arXiv:2106.08664 [hep-ph]]}.

\bibitem{Chawdhry:2021mkw}
H.~A.~Chawdhry, M.~Czakon, A.~Mitov and R.~Poncelet,
\href{https://link.springer.com/article/10.1007/JHEP07(2021)164}{JHEP \textbf{07} (2021), 164}
\href{https://arxiv.org/abs/2103.04319}{[arXiv:2103.04319 [hep-ph]]}.

\bibitem{Agarwal:2021grm}
B.~Agarwal, F.~Buccioni, A.~von Manteuffel and L.~Tancredi,
\href{https://link.springer.com/article/10.1007/JHEP04(2021)201}{JHEP \textbf{04} (2021), 201}
\href{https://arxiv.org/abs/2102.01820}{[arXiv:2102.01820 [hep-ph]]}.

\bibitem{Chawdhry:2020for}
H.~A.~Chawdhry, M.~Czakon, A.~Mitov and R.~Poncelet,
\href{https://link.springer.com/article/10.1007/JHEP06(2021)150}{JHEP \textbf{06} (2021), 150}
\href{https://arxiv.org/abs/2012.13553}{[arXiv:2012.13553 [hep-ph]]}.

\bibitem{Abreu:2020cwb}
S.~Abreu, B.~Page, E.~Pascual and V.~Sotnikov,
\href{https://link.springer.com/article/10.1007/JHEP01(2021)078}{JHEP \textbf{01} (2021), 078}
\href{https://arxiv.org/abs/2010.15834}{[arXiv:2010.15834 [hep-ph]]}.

\bibitem{Abreu:2021oya}
S.~Abreu, F.~Febres Cordero, H.~Ita, B.~Page and V.~Sotnikov,
\href{https://link.springer.com/article/10.1007/JHEP07(2021)095}{JHEP \textbf{07} (2021), 095}
\href{https://arxiv.org/abs/2102.13609}{[arXiv:2102.13609 [hep-ph]]}.

\bibitem{Abreu:2019odu}
S.~Abreu, J.~Dormans, F.~Febres Cordero, H.~Ita, B.~Page and V.~Sotnikov,
\href{https://link.springer.com/article/10.1007/JHEP05(2019)084}{JHEP \textbf{05} (2019), 084}
\href{https://arxiv.org/abs/1904.00945}{[arXiv:1904.00945 [hep-ph]]}.

\bibitem{Abreu:2018zmy}
S.~Abreu, J.~Dormans, F.~Febres Cordero, H.~Ita and B.~Page,
\href{https://journals.aps.org/prl/abstract/10.1103/PhysRevLett.122.082002}{Phys. Rev. Lett. \textbf{122} (2019) no.8, 082002}
\href{https://arxiv.org/abs/1812.04586}{[arXiv:1812.04586 [hep-ph]]}.

\bibitem{Abreu:2023bdp}
S.~Abreu, G.~De Laurentis, H.~Ita, M.~Klinkert, B.~Page and V.~Sotnikov,
\href{https://arxiv.org/abs/2305.17056}{[arXiv:2305.17056 [hep-ph]]}.

\bibitem{Agarwal:2021vdh}
B.~Agarwal, F.~Buccioni, A.~von Manteuffel and L.~Tancredi,
\href{https://journals.aps.org/prl/abstract/10.1103/PhysRevLett.127.262001}{Phys. Rev. Lett. \textbf{127} (2021) no.26, 262001}
\href{https://arxiv.org/abs/2105.04585}{[arXiv:2105.04585 [hep-ph]]}.

\bibitem{Peraro:2016wsq}
T.~Peraro,
\href{https://link.springer.com/article/10.1007/JHEP12(2016)030}{JHEP \textbf{12} (2016), 030}
\href{https://arxiv.org/abs/1608.01902}{[arXiv:1608.01902 [hep-ph]]}.

\bibitem{Pozzorini:2022ohr}
S.~Pozzorini, N.~Sch\"ar and M.~F.~Zoller,
\href{https://link.springer.com/article/10.1007/JHEP05(2022)161}{JHEP \textbf{05} (2022), 161}
\href{https://arxiv.org/abs/2201.11615}{[arXiv:2201.11615 [hep-ph]]}.

\bibitem{DE1}
A.~V.~Kotikov,
\href{https://doi.org/10.1016/0370-2693(91)90413-K}{Phys. Lett. B \textbf{254} (1991), 158-164}.

\bibitem{DE2}
A.~V.~Kotikov,
\href{https://doi.org/10.1016/0370-2693(91)90536-Y}{Phys. Lett. B \textbf{267} (1991), 123-127}
[erratum: \href{https://doi.org/10.1016/0370-2693(92)91582-T}{Phys. Lett. B \textbf{295} (1992), 409-409}].

\bibitem{Gehrmann:1999as}
T.~Gehrmann and E.~Remiddi,
\href{https://www.sciencedirect.com/science/article/abs/pii/S0550321300002236?via%3Dihub}{Nucl. Phys. B \textbf{580} (2000), 485-518}
\href{https://arxiv.org/abs/hep-ph/9912329}{[arXiv:hep-ph/9912329 [hep-ph]]}.


\bibitem{Henn:2013pwa}
J.~M.~Henn,
\href{https://journals.aps.org/prl/abstract/10.1103/PhysRevLett.110.251601}{Phys. Rev. Lett. \textbf{110} (2013), 251601}
\href{https://arxiv.org/abs/1304.1806}{[arXiv:1304.1806 [hep-th]]}.

\bibitem{Papadopoulos:2014lla}
C.~G.~Papadopoulos,
\href{https://link.springer.com/article/10.1007/JHEP07(2014)088}{JHEP \textbf{07} (2014), 088}
\href{https://arxiv.org/abs/1401.6057}{[arXiv:1401.6057 [hep-ph]]}.

\bibitem{Papadopoulos:2015jft}
C.~G.~Papadopoulos, D.~Tommasini and C.~Wever,
\href{https://link.springer.com/article/10.1007/JHEP04(2016)078}{JHEP \textbf{04} (2016), 078}
\href{https://arxiv.org/abs/1511.09404}{[arXiv:1511.09404 [hep-ph]]}.

\bibitem{Liu:2017jxz}
X.~Liu, Y.~Q.~Ma and C.~Y.~Wang,
\href{https://www.sciencedirect.com/science/article/pii/S037026931830128X?via%3Dihub}{Phys. Lett. B \textbf{779} (2018), 353-357}
\href{https://arxiv.org/abs/1711.09572}{[arXiv:1711.09572 [hep-ph]]}.

\bibitem{Gehrmann:2018yef}
T.~Gehrmann, J.~M.~Henn and N.~A.~Lo Presti,
\href{https://link.springer.com/article/10.1007/JHEP10(2018)103}{JHEP \textbf{10} (2018), 103}
\href{https://arxiv.org/abs/1807.09812}{[arXiv:1807.09812 [hep-ph]]}.

\bibitem{Chicherin:2018old}
D.~Chicherin, T.~Gehrmann, J.~M.~Henn, P.~Wasser, Y.~Zhang and S.~Zoia,
\href{https://journals.aps.org/prl/abstract/10.1103/PhysRevLett.123.041603}{Phys. Rev. Lett. \textbf{123} (2019) no.4, 041603}
\href{https://arxiv.org/abs/1812.11160}{[arXiv:1812.11160 [hep-ph]]}.

\bibitem{Chicherin:2020oor}
D.~Chicherin and V.~Sotnikov,
\href{https://link.springer.com/article/10.1007/JHEP12(2020)167}{JHEP \textbf{20} (2020), 167}
\href{https://arxiv.org/abs/2009.07803}{[arXiv:2009.07803 [hep-ph]]}.

\bibitem{Abreu:2020jxa}
S.~Abreu, H.~Ita, F.~Moriello, B.~Page, W.~Tschernow and M.~Zeng,
\href{https://link.springer.com/article/10.1007/JHEP11(2020)117}{JHEP \textbf{11} (2020), 117}
\href{https://arxiv.org/abs/2005.04195}{[arXiv:2005.04195 [hep-ph]]}.

\bibitem{Canko:2020ylt}
D.~D.~Canko, C.~G.~Papadopoulos and N.~Syrrakos,
\href{https://link.springer.com/article/10.1007/JHEP01(2021)199}{JHEP \textbf{01} (2021), 199}
\href{https://arxiv.org/abs/2009.13917}{[arXiv:2009.13917 [hep-ph]]}.

\bibitem{Abreu:2021smk}
S.~Abreu, H.~Ita, B.~Page and W.~Tschernow,
\href{https://link.springer.com/article/10.1007/JHEP03(2022)182}{JHEP \textbf{03} (2022), 182}
\href{https://arxiv.org/abs/2107.14180}{[arXiv:2107.14180 [hep-ph]]}.

\bibitem{Kardos:2022tpo}
A.~Kardos, C.~G.~Papadopoulos, A.~V.~Smirnov, N.~Syrrakos and C.~Wever,
\href{https://link.springer.com/article/10.1007/JHEP05(2022)033}{JHEP \textbf{05} (2022), 033}
\href{https://arxiv.org/abs/2201.07509}{[arXiv:2201.07509 [hep-ph]]}.

\bibitem{Chicherin:2021dyp}
D.~Chicherin, V.~Sotnikov and S.~Zoia,
\href{https://link.springer.com/article/10.1007/JHEP01(2022)096}{JHEP \textbf{01} (2022), 096}
\href{https://arxiv.org/abs/2110.10111}{[arXiv:2110.10111 [hep-ph]]}.

\bibitem{Abreu:2023rco}
S.~Abreu, D.~Chicherin, H.~Ita, B.~Page, V.~Sotnikov, W.~Tschernow and S.~Zoia,
\href{https://arxiv.org/abs/2306.15431}{[arXiv:2306.15431 [hep-ph]]}.

\bibitem{SecDec}
T.~Binoth and G.~Heinrich,
\href{https://doi.org/10.1016/S0550-3213(00)00429-6}{Nucl. Phys. B \textbf{585} (2000), 741-759}
\href{https://arxiv.org/abs/hep-ph/0004013}{\tt [hep-ph/0004013]}.

\bibitem{Borinsky:2020rqs}
M.~Borinsky,
\href{https://arxiv.org/abs/2008.12310}{[arXiv:2008.12310 [math-ph]]}.

\bibitem{Moriello:2019yhu}
F.~Moriello,
\href{https://link.springer.com/article/10.1007/JHEP01(2020)150}{JHEP \textbf{01} (2020), 150}
\href{https://arxiv.org/abs/1907.13234}{[arXiv:1907.13234 [hep-ph]]}.


\bibitem{Kallweit:2020gcp}
S.~Kallweit, V.~Sotnikov and M.~Wiesemann,
\href{https://www.sciencedirect.com/science/article/pii/S0370269320308169?via%3Dihub}{Phys. Lett. B \textbf{812} (2021), 136013}
\href{https://arxiv.org/abs/2010.04681}{[arXiv:2010.04681 [hep-ph]]}.

\bibitem{Chawdhry:2021hkp}
H.~A.~Chawdhry, M.~Czakon, A.~Mitov and R.~Poncelet,
\href{https://link.springer.com/article/10.1007/JHEP09(2021)093}{JHEP \textbf{09} (2021), 093}
\href{https://arxiv.org/abs/2105.06940}{[arXiv:2105.06940 [hep-ph]]}.

\bibitem{Czakon:2021mjy}
M.~Czakon, A.~Mitov and R.~Poncelet,
\href{https://journals.aps.org/prl/abstract/10.1103/PhysRevLett.127.152001}{Phys. Rev. Lett. \textbf{127} (2021) no.15, 152001}
[erratum: \href{https://journals.aps.org/prl/abstract/10.1103/PhysRevLett.129.119901}{Phys. Rev. Lett. \textbf{129} (2022) no.11, 119901}]
\href{https://arxiv.org/abs/2106.05331}{[arXiv:2106.05331 [hep-ph]]}.

\bibitem{Hartanto:2022qhh}
H.~B.~Hartanto, R.~Poncelet, A.~Popescu and S.~Zoia,
\href{https://journals.aps.org/prd/abstract/10.1103/PhysRevD.106.074016}{Phys. Rev. D \textbf{106} (2022) no.7, 074016}
\href{https://arxiv.org/abs/2205.01687}{[arXiv:2205.01687 [hep-ph]]}.

\bibitem{Badger:2023mgf}
S.~Badger, M.~Czakon, H.~B.~Hartanto, R.~Moodie, T.~Peraro, R.~Poncelet and S.~Zoia,
\href{https://arxiv.org/abs/2304.06682}{[arXiv:2304.06682 [hep-ph]]}.

\bibitem{tHooft:1973alw}
G.~'t Hooft, 
\href{https://www.sciencedirect.com/science/article/pii/0550321374901540?via%3Dihub}{Nucl.Phys.B 72 (1974) 461}.


\bibitem{Kanaki:2000ey}
A.~Kanaki and C.~G.~Papadopoulos,
\href{https://www.sciencedirect.com/science/article/pii/S001046550000151X?via%3Dihub}{Comput.Phys.Commun. 132 (2000) 306-315}
\href{https://arxiv.org/abs/hep-ph/0002082}{[arXiv:hep-ph/0002082]}.

\bibitem{Cafarella:2007pc}
A.~Cafarella, C.~G.~Papadopoulos and M.~Worek,
\href{https://www.sciencedirect.com/science/article/pii/S0010465509001465?via%3Dihub}{Comput.Phys.Commun. 180 (2009) 1941-1955}
\href{https://arxiv.org/abs/0710.2427}{[arXiv:0710.2427 [hep-ph]]}.


\bibitem{Bevilacqua:2021dym}
G.~Bevilacqua, D.~D.~Canko, A.~Kardos and C.~G.~Papadopoulos,
\href{https://iopscience.iop.org/article/10.1088/1742-6596/2105/1/012010}{J. Phys. Conf. Ser. \textbf{2105} (2021) no.5, 012010}.


\bibitem{vanHameren:2009dr}
A.~van Hameren, C.~G.~Papadopoulos and R.~Pittau,
\href{https://iopscience.iop.org/article/10.1088/1126-6708/2009/09/106}{JHEP \textbf{09} (2009), 106}
\href{https://arxiv.org/abs/0903.4665}{[arXiv:0903.4665 [hep-ph]]}.

\bibitem{Bevilacqua:2011xh}
G.~Bevilacqua, M.~Czakon, M.~V.~Garzelli, A.~van Hameren, A.~Kardos, C.~G.~Papadopoulos, R.~Pittau and M.~Worek,
\href{https://www.sciencedirect.com/science/article/abs/pii/S0010465512003761?via%3Dihub}{Comput. Phys. Commun. \textbf{184} (2013), 986-997}
\href{https://arxiv.org/abs/1110.1499}{[arXiv:1110.1499 [hep-ph]]}.


\bibitem{Hahn:2000kx}
T.~Hahn,
\href{https://www.sciencedirect.com/science/article/abs/pii/S0010465501002909?via%3Dihub}{Comput. Phys. Commun. \textbf{140} (2001), 418-431}
\href{https://arxiv.org/abs/hep-ph/0012260}{[arXiv:hep-ph/0012260 [hep-ph]]}.

\bibitem{Shtabovenko:2020gxv}
V.~Shtabovenko, R.~Mertig and F.~Orellana,
\href{https://www.sciencedirect.com/science/article/abs/pii/S001046552030223X?via%3Dihub}{Comput. Phys. Commun. \textbf{256} (2020), 107478}
\href{https://arxiv.org/abs/2001.04407}{[arXiv:2001.04407 [hep-ph]]}.

\bibitem{new_paper}
G.~Bevilacqua, D.~D.~Canko and C.~G.~Papadopoulos, work in progress.


\bibitem{Chetyrkin:1981qh}
K.~G.~Chetyrkin and F.~V.~Tkachov,
\href{https://www.sciencedirect.com/science/article/abs/pii/0550321381901991?via%3Dihub}{Nucl. Phys. B \textbf{192} (1981), 159-204}.

\bibitem{Tkachov:1981wb}
F.~V.~Tkachov,
\href{https://www.sciencedirect.com/science/article/abs/pii/0370269381902884?via%3Dihub}{Phys. Lett. B \textbf{100} (1981), 65-68}.

\bibitem{Laporta:2000dsw}
S.~Laporta,
\href{https://www.worldscientific.com/doi/abs/10.1142/S0217751X00002159}{Int. J. Mod. Phys. A \textbf{15} (2000), 5087-5159}
\href{https://arxiv.org/abs/hep-ph/0102033}{[arXiv:hep-ph/0102033 [hep-ph]]}.


\bibitem{Smirnov:2019qkx}
A.~V.~Smirnov and F.~S.~Chuharev,
\href{https://www.sciencedirect.com/science/article/abs/pii/S0010465519302644?via%3Dihub}{Comput. Phys. Commun. \textbf{247 } (2020), 106877}
\href{https://arxiv.org/abs/1901.07808}{[arXiv:1901.07808 [hep-ph]]}.

\bibitem{Klappert:2020nbg}
J.~Klappert, F.~Lange, P.~Maierh\"ofer and J.~Usovitsch,
\href{https://www.sciencedirect.com/science/article/abs/pii/S0010465521001363?via%3Dihub}{Comput. Phys. Commun. \textbf{266} (2021), 108024}
\href{https://arxiv.org/abs/2008.06494}{[arXiv:2008.06494 [hep-ph]]}.


\bibitem{Sotnikov:2019onv}
V.~Sotnikov,
PhD Thesis (2019),
\href{https://freidok.uni-freiburg.de/data/151540}{DOI: 10.6094/UNIFR/151540}.

\bibitem{Ossola:2006us}
G.~Ossola, C.~G.~Papadopoulos and R.~Pittau,
\href{https://www.sciencedirect.com/science/article/abs/pii/S0550321306009138?via%3Dihub}{Nucl. Phys. B \textbf{763} (2007), 147-169}
\href{https://arxiv.org/abs/hep-ph/0609007}{[arXiv:hep-ph/0609007 [hep-ph]]}.

\bibitem{Ossola:2007ax}
G.~Ossola, C.~G.~Papadopoulos and R.~Pittau,
\href{https://iopscience.iop.org/article/10.1088/1126-6708/2008/03/042}{JHEP \textbf{03} (2008), 042}
\href{https://arxiv.org/abs/0711.3596}{[arXiv:0711.3596 [hep-ph]]}.


\bibitem{Pozzorini:2020hkx}
S.~Pozzorini, H.~Zhang and M.~F.~Zoller,
\href{https://link.springer.com/article/10.1007/JHEP05(2020)077}{JHEP \textbf{05} (2020), 077}
\href{https://arxiv.org/abs/2001.11388}{[arXiv:2001.11388 [hep-ph]]}.

\bibitem{Lang:2020nnl}
J.~N.~Lang, S.~Pozzorini, H.~Zhang and M.~F.~Zoller,
\href{https://link.springer.com/article/10.1007/JHEP10(2020)016}{JHEP \textbf{10} (2020), 016}
\href{https://arxiv.org/abs/2007.03713}{[arXiv:2007.03713 [hep-ph]]}.

\bibitem{Lang:2021hnw}
J.~N.~Lang, S.~Pozzorini, H.~Zhang and M.~F.~Zoller,
\href{https://link.springer.com/article/10.1007/JHEP01(2022)105}{JHEP \textbf{01} (2022), 105}
\href{https://arxiv.org/abs/2107.10288}{[arXiv:2107.10288 [hep-ph]]}.


\end{thebibliography}
\end{document}